\documentclass[twocolumn,aps,prl,preprintnumbers]{revtex4}
\usepackage{}
\usepackage{bbm}
\usepackage[latin9]{inputenc}
\usepackage{amsmath}
\usepackage{amssymb}
\usepackage{graphicx}
\usepackage{mathrsfs}
\usepackage{amsfonts}
\usepackage{amsthm}
\usepackage{color}
\usepackage{txfonts}
\usepackage[colorlinks=true,citecolor=blue,linkcolor=blue,urlcolor=blue,anchorcolor=blue]{hyperref}%
\hypersetup{colorlinks=true,citecolor=blue,linkcolor=blue,urlcolor=blue}
\providecommand{\U}[1]{\protect\rule{.1in}{.1in}}
\makeatletter
\@ifundefined{textcolor}{}
{
\definecolor{BLACK}{gray}{0}
\definecolor{WHITE}{gray}{1}
\definecolor{RED}{rgb}{1,0,0}
\definecolor{GREEN}{rgb}{0,1,0}
\definecolor{BLUE}{rgb}{0,0,1}
\definecolor{CYAN}{cmyk}{1,0,0,0}
\definecolor{MAGENTA}{cmyk}{0,1,0,0}
\definecolor{YELLOW}{cmyk}{0,0,1,0}
}
\makeatother
\begin{document}
\title{Observation of symmetry protected zero modes in topolectrical circuits}
\author{Huanhuan Yang$^{1}$}
\author{Z.-X. Li$^{1}$}
\author{Yuanyuan Liu$^{1}$}
\author{Yunshan Cao$^{1}$}
\author{Peng Yan$^{1}$}
\email[Corresponding author: ]{yan@uestc.edu.cn}
\affiliation{$^{1}$School of Electronic Science and Engineering and State Key Laboratory of Electronic Thin Films and Integrated Devices, University of
Electronic Science and Technology of China, Chengdu 610054, China}

\begin{abstract}
Higher-order topological insulators are a new class of topological phases of matter, originally conceived for electrons in solids. It has been suggested that $\mathbb{Z}_N$ Berry phase (Berry phase quantized into $2\pi/N$) is a useful tool to characterize the symmetry protected topological states, while the experimental evidence is still elusive. Recently, topolectrical circuits have emerged as a simple yet very powerful platform for studying topological physics that are challenging to realize in condensed matter systems. Here, we present the first experimental observation of second-order corner states characterized by $\mathbb{Z}_3$ Berry phase in topolectrical circuits. We demonstrate theoretically and experimentally that the localized second-order topological states are protected by a generalized chiral symmetry of tripartite lattices, and they are pinned to ``zero energy". By introducing extra capacitors within sublattices in the circuit, we are able to examine the robustness of the zero modes against both chiral-symmetry conserving and breaking disturbances. Our work paves the way for testing exotic topological band theory by electrical-circuit experiments.
\end{abstract}

\maketitle
\emph{Introduction.} Since the theoretical prediction in electron system \cite{Benalcazar2017,Bernevig2017,Song2017,Langbehn2017,Schindler2018,Ezawa2018_1,Khalaf2018,Geier2018,Queiro2019,Benalcazar2019,Peterson2018,Peterson2020}, higher-order topological insulators (HOTIs) have attracted considerable attention by the broad community of photonics \cite{Xie2018,Noh2018,Hassan2019,Mittal2019,Chen2019,Xie2019,Ota2019,ZhangL2019,LiM2019}, acoustics \cite{Xue2019,Ni2019,Xue2019_2,He2019,ZhangX2019,ChenZ2019,ZhangZ2019}, mechanics \cite{Serra-Garcia2018,Fan2019}, and very recently, spintronics \cite{Li2019_1,Li2019_2}. The extensive research enthusiasm is sparked by the peculiar properties of HOTIs at device corners or hinges that are of fundamental interest in terms of extended bulk-boundary correspondence, and are of promising applications in memory, computing, and imaging \cite{Zhang2019}. The Berry phase offers a powerful means to describe the topological feature of band structures \cite{Zak1989}. It has been suggested that $\mathbb{Z}_N$ Berry phase (quantized to $2\pi/N$) is a useful tool to characterize the symmetry-protected higher-order topological phase \cite{Kariyado2018,Araki2019,Kudo2019}. It is argued that the generalized chiral symmetry (sublattice symmetry) in the tripartite \cite{Kempkes2019} and sexpartite \cite{Li2019_2} lattices can protect the topological ``zero-energy" corner states that are particularly robust against disorder and defects. The experimental evidence of the topological stability of the zero mode, however, is still elusive. The major challenge is that it is quite difficult to selectively introduce chiral-symmetry conserving or breaking disturbances in condensed matter experiments.

Recently, topolectrical circuits have emerged as a simple yet formidable platform for topological physics \cite{Lee2018,Helbig2019,Lu2019Y,Imhof2018,Nejad2019,Serra-Garcia2019,Bao2019}. In this Rapid Communication, we report the first experimental observation of second-order corner states characterized by $\mathbb{Z}_3$ Berry phase in topolectrical circuits. We demonstrate theoretically and experimentally that the localized second-order topological states are pinned to ``zero energy". By introducing external capacitors connecting different sublattices in the circuit, we examine the robustness of the zero modes against both chiral-symmetry conserving and breaking disturbances. Through two-point impedance measurements, we observe that the zero mode is robust if and only if the disturbing connection via capacitors conserves the generalized chiral symmetry. Our findings therefore open the door for testing exotic topological band model in electrical-circuit experiments.
\begin{figure}[htbp!]
  \centering
  \includegraphics[width=0.48\textwidth]{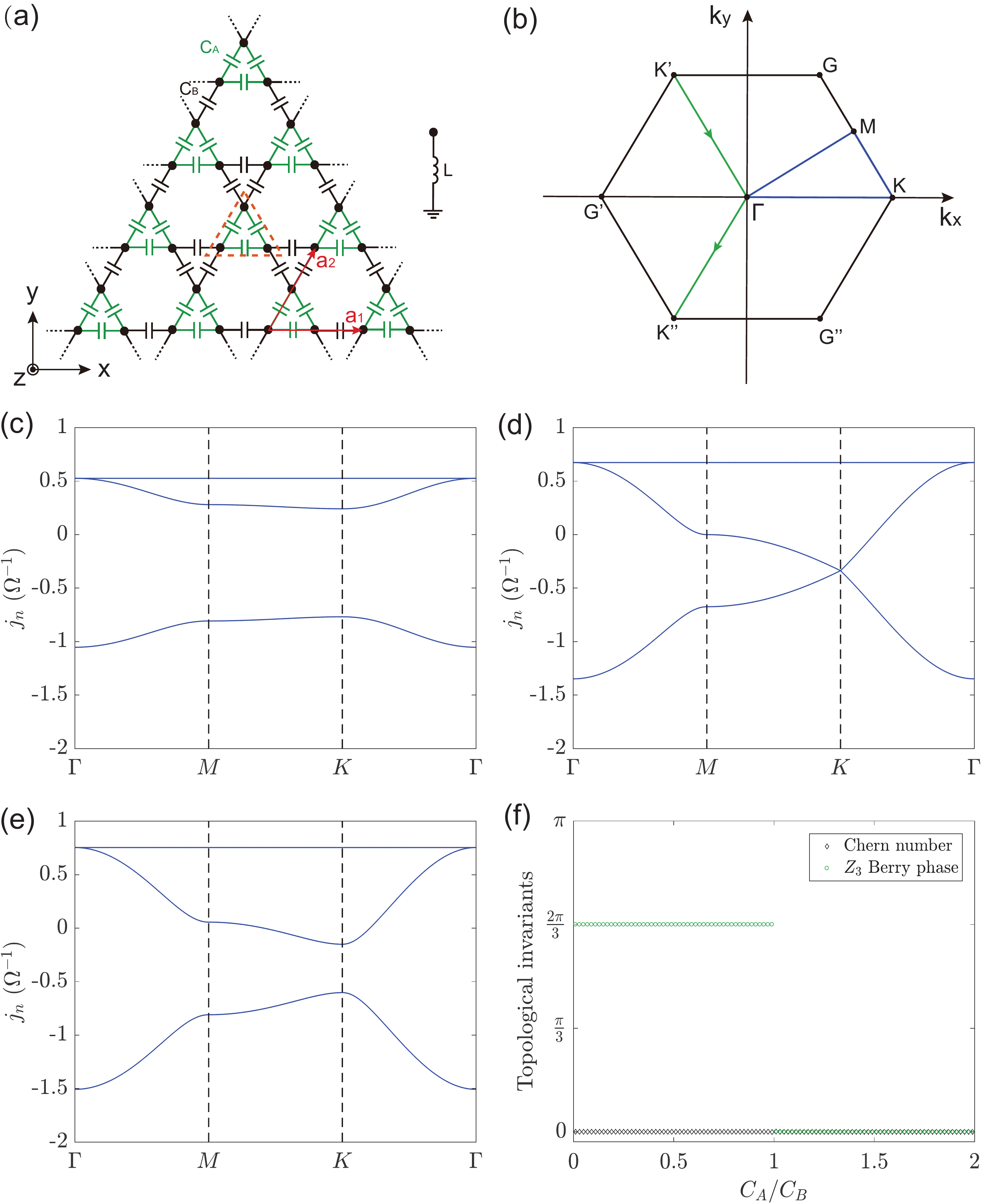}\\
   \caption{(a) Illustration of an extended breathing kagome electrical circuit, consisting of two types of capacitors ($C_A,C_B$). The black dots are nodes grounded through an inductor ($L$). (b) The first Brillouin zone. (c)-(e) Band structure for different capacitance ratio: $C_A/C_B=0.22$ (c), $C_A/C_B=1$ (d), and $C_A/C_B=1.5$ (e). (f) $\mathbb{Z}_3$ Berry phase and Chern number as a function of $C_A/C_B$.}\label{model}
\end{figure}

\emph{Kagome electrical circuit.} We consider non-dissipative linear electric circuits made of capacitors and inductors. Labelling the nodes of a circuit by $a=1,2,...$, the response of the circuit at frequency $\omega$ follows Kirchhoff's law:
\begin{equation}
I_a(\omega)=\sum_bJ_{ab}(\omega)V_b(\omega),
\end{equation}
with
\begin{equation}
J_{ab}(\omega)=i\omega[C_{ab}+\delta_{ab}(\sum_nC_{an}-\frac{1}{\omega^2L_a})],
\end{equation}
where $I_a$ is the external current flow into node $a$, $J_{ab}$($\omega$) is circuit Laplacian, $V_b$ is the voltage at node $b$, $C_{ab}$ is the capacitor between nodes $a$ and $b$, $L_a$ is the grounded inductance at node $a$, and the sum is taken over all neighboring nodes. Analogous to the tight-binding Hamiltonian formalism in condensed matter physics, $J_{ab}$($\omega$) can be expressed as $i \mathcal{H}_{ab}(\omega)$ with $\mathcal{H}$ being a Hermitian matrix, making the following analysis convenient.

Figure \ref{model}(a) shows the structure for an infinite breathing kagome lattice of circuit: the dashed orange triangle represents the unit cell including three sites (nodes), with the first Brillouin zone plotted in Fig. \ref{model}(b). $\textbf{a}_{1}=d\hat{x}$ and $\textbf{a}_{2}=\frac{1}{2}d\hat{x}+\frac{\sqrt{3}}{2}d\hat{y}$ are the two basis vectors with $d$ the lattice constant. The Hamiltonian $\mathcal{H}$ of the electrical circuit can be expressed as
\begin{equation}\label{Eq1}
 \mathcal {H}=\left(
 \begin{matrix}
   Q_{0} & Q_{1} & Q_{2}\\
   Q_{1}^{*} & Q_{0} & Q_{3} \\
   Q_{2}^{*} & Q_{3}^{*} & Q_{0}
  \end{matrix}
  \right),
\end{equation}
with the matrix elements:
\begin{equation}\label{Eq2}
\begin{aligned}
Q_{0}&=\omega[2(C_{A}+C_{B})-\frac{1}{\omega^{2}L}], \\
Q_{1}&=-\omega[C_{A}+C_{B}e^{-i(k_{x}d/2+\sqrt{3}k_{y}d/2)}], \\
Q_{2}&=-\omega[C_{A}+C_{B}e^{-ik_{x}d}],\\
Q_{3}&=-\omega[C_{A}+C_{B}e^{i(-k_{x}d/2+\sqrt{3}k_{y}d/2)}].\\
\end{aligned}
\end{equation}
Figures \ref{model}(c)-\ref{model}(e) show the band structures for different capacitance ratios $C_{A}/C_{B}$, which is gapless when $C_{A}/C_{B}=1$ [see Fig. \ref{model}(d)]. However, the gap opens at $K$ point if $C_{A}\neq C_{B}$ [see Figs. \ref{model}(c) and \ref{model}(e)], leading to an insulating phase.

\emph{Generalized chiral symmetry and $\mathbb{Z}_{3}$ Berry phase.} For any insulator with translational symmetry, the gauge-invariant Chern number \cite{Avron1983,Wang2017}:
\begin{equation}\label{Eq5}
   \mathcal{C}=\frac{i}{2\pi}\int\!\!\!\int_{\text{BZ}}dk_{x}dk_{y}\text{Tr}\big[P( \frac{\partial P}{\partial k_{x}}\frac{\partial P}{\partial k_{y}}- \frac{\partial P}{\partial k_{y}}\frac{\partial P}{\partial k_{x}})\big]
\end{equation}
can be used for determining the first-order topological phase, where $P$ is the projection matrix $P(\textbf{k})=\phi (\textbf{k})\phi (\textbf{k})^{\dag}$, with $\phi (\textbf{k})$ being the normalized eigenstate of \eqref{Eq1} in any band, i.e., the 1st, the 2nd, or the 3rd band, and the integral is over the first Brillouin zone (BZ). However, to judge if the system allows higher-order topological phases, one needs a different topological invariant. To this end, we consider the $\mathbb{Z}_{3}$ Berry phase:
\begin{equation}\label{Eq6}
 \mathcal{\theta}=\int_{L_{1}}\text{Tr}[\textbf{A}(\textbf{k})]\cdot d\textbf{k}\ \  (\text{mod}\ 2\pi),
\end{equation}
where $\textbf{A}(\textbf{k})$ is the Berry connection:
\begin{equation}\label{Eq7}
 \textbf{A}(\textbf{k})=i\Psi^{\dag}(\textbf{k})\frac{\partial}{\partial\textbf{k}}\Psi(\textbf{k}).
\end{equation}
Here, $\Psi(\textbf{k})$ is the eigenvector of \eqref{Eq1} for the lowest  (the 1st) band. $L_{1}$ is an integral path in BZ: $K^{\prime}\rightarrow \Gamma\rightarrow K^{\prime\prime}$; see the green line segment in Fig. \ref{model}(b). It is noted that there are other two equivalent paths for calculating $\mathcal{\theta}$ ($L_{2}: K^{\prime\prime}\rightarrow \Gamma\rightarrow K$, $L_{3}: K\rightarrow \Gamma\rightarrow K^{\prime}$) because of the $C_{3}$ symmetry (the three high-symmetry points $K$, $K^{\prime}$, and $K^{\prime\prime}$ in the BZ are equivalent). Furthermore, it is obvious that the integral along $L_{1}+L_{2}+L_{3}$ vanishes. The $\mathbb{Z}_{3}$ Berry phase thus must be quantized as $\theta=\frac{2n\pi}{3}\ $ $(n=0,1,2)$ \cite{Zak1989,Kariyado2018,Araki2019}. By simultaneously evaluating the Chern number and $\mathbb{Z}_{3}$ Berry phase, one can completely determine the topological phase allowed in the electrical circuit.

We note that the diagonal element $Q_{0}$ becomes zero when $\omega$ matches the resonant condition, i.e., $\omega=\omega_{c}=1/\sqrt{2L(C_{A}+C_{B})}$, which is the ``zero-energy" of the electrical-circuit system. Although the breathing kagome lattice is not a bipartite lattice, we prove that the ``zero-energy'' modes (corner states) are protected by a generalized chiral symmetry, see Supplemental Material (SM) \cite{SM}.

Figure \ref{model}(f) shows the dependence of the Chern number $\mathcal{C}$ and $\mathbb{Z}_{3}$ Berry phase $\theta$ on the ratio $C_{A}/C_{B}$. One finds that $\theta$ is quantized to $2\pi/3$ for $C_{A}/C_{B}<1$, and to $0$ for $C_{A}/C_{B}>1$, with $C_{A}/C_{B}=1$ being the phase transition point. Meanwhile, the Chern number $\mathcal{C}$ vanishes for all $C_{A}/C_{B}$, indicating that the system does not support any topological chiral edge phase (first-order topological insulator state). We thus conclude that the system allows only two topologically distinct phases for $C_{A}/C_{B}<1$, and for $C_{A}/C_{B}>1$.
\begin{figure}[htb]
  \centering
  \includegraphics[width=0.48\textwidth]{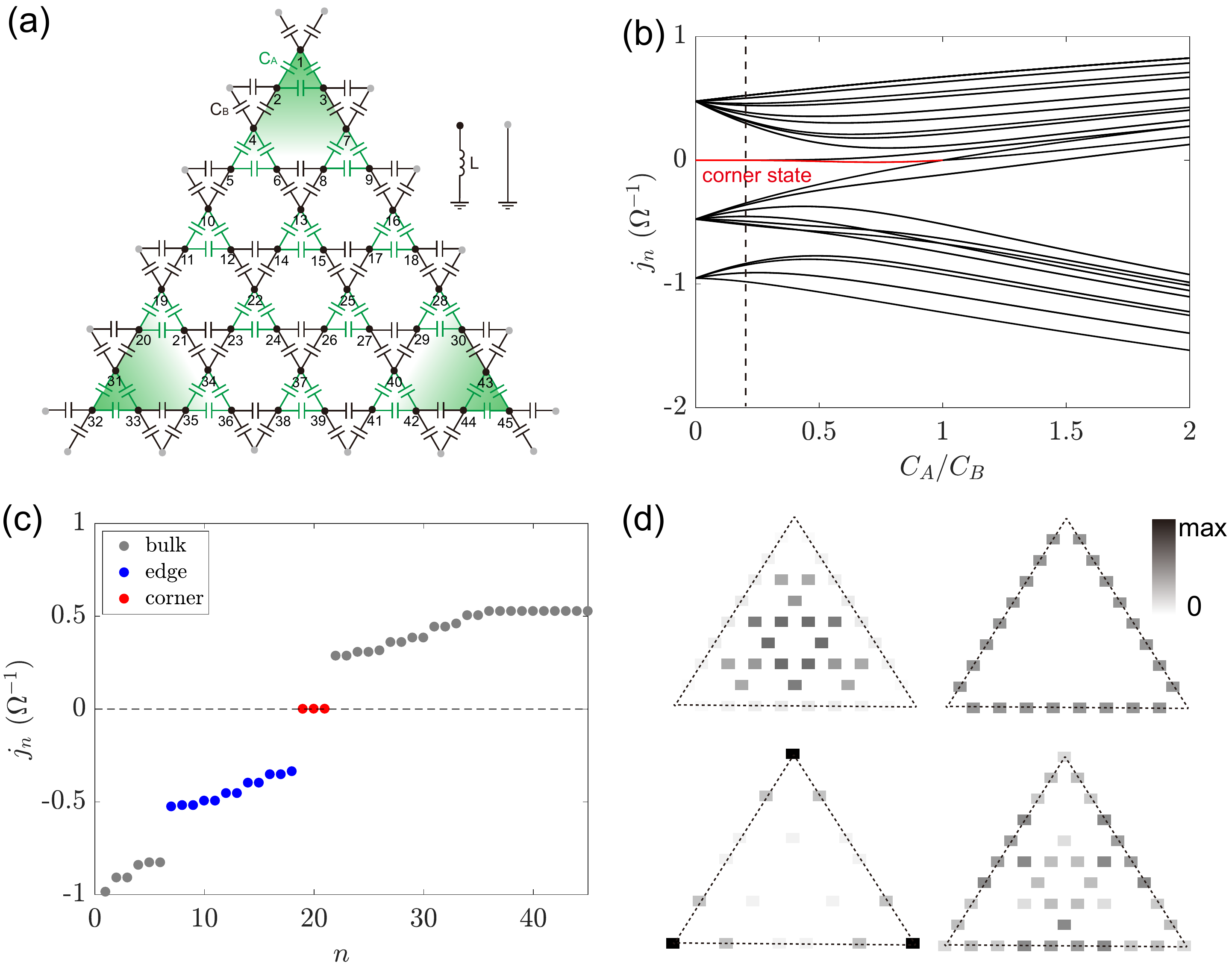}\\
  \caption{(a) Schematic plot of a finite-size circuit with the gray dots being grounded directly. (b) The admittance spectrum at the resonant condition under different ratios $C_A/C_B$ with the red segment denoting the corner-state phase. (c) Admittances for $C_A/C_B$=0.22 [dashed black line in (b)]. The gray, blue, and red dots represent the bulk, edge, and corner states, respectively. (d) Spatial distribution of the bulk, edge, corner, and bulk modes (from left to right and top to bottom) with $j_n=-0.9858$, $-0.3367$, $1.659\times10^{-4}$, $0.3147$ $\Omega^{-1}$, respectively.}\label{ES}
\end{figure}

\emph{Zero modes in triangle-shape electrical circuit.} Considering a finite-size breathing kagome circuit of triangle shape with $\mathcal{N}=45$ nodes, as shown in Fig. \ref{ES}(a), the circuit Laplacian $J(\omega)$ reads:
\begin{equation}\label{Eq8}
J(\omega)=\left(
\begin{array}{cccccc}
J_0  & -J_A  & -J_A &  0   &   0   & \ldots\\
-J_A & J_0   & -J_A & -J_B &   0   & \ldots \\
-J_A & -J_A  & J_0  &  0   &  0 & \ldots\\
0    &  -J_B & 0    &  J_0 &  -J_A    & \ldots\\
0    & 0     & 0 &  -J_A   &  J_0  & \ldots\\
\vdots  & \vdots  & \vdots &  \vdots   &  \vdots  & \ddots \\
\end{array}
\right)_{\mathcal{N}\times \mathcal{N}},
\end{equation}
where $J_0=2i\omega (C_A+C_B)+1/(i\omega L)$, $J_A=i\omega C_A$, and $J_B=i\omega C_B$. By diagonalizing \eqref{Eq8}, we obtain both eigenvalues $j_n$ (admittances) and eigenfunctions $\psi_n$, with $n=1,2,...,\mathcal{N}$. In Fig. \ref{ES}(b), we calculate the admittance spectrum for $C_A/C_B$ ranging from 0 to 2. The zero mode appears when $0<C_A/C_B<1$ (the red line segment), corresponding to the HOTI phase, among which a kind of bound states in the continuum emerges for $0.5<C_A/C_B<1$ \cite{Benalcazar20192}. A representative example at $C_A/C_B=0.22$ is shown in Fig. \ref{ES}(c). We identify three kinds of eigenmodes by gray, blue, and red dots, respectively, with the spatial distribution of the wavefunctions being plotted in Fig. \ref{ES}(d). It clearly shows that the three degenerate modes (red dots) inside the band gap are corner states, while the blue and gray dots represent the edge and bulk states, respectively. Interestingly, the $\mathbb{Z}_{3}$ Berry phase theory indicates that corner states can actually survive until $C_{A}/C_{B}=1$ rather than $1/2$ as claimed in Ref. \cite{Ezawa2018}. We confirm this point by experimental measurements.
\begin{figure}
  \centering
  \includegraphics[width=0.48\textwidth]{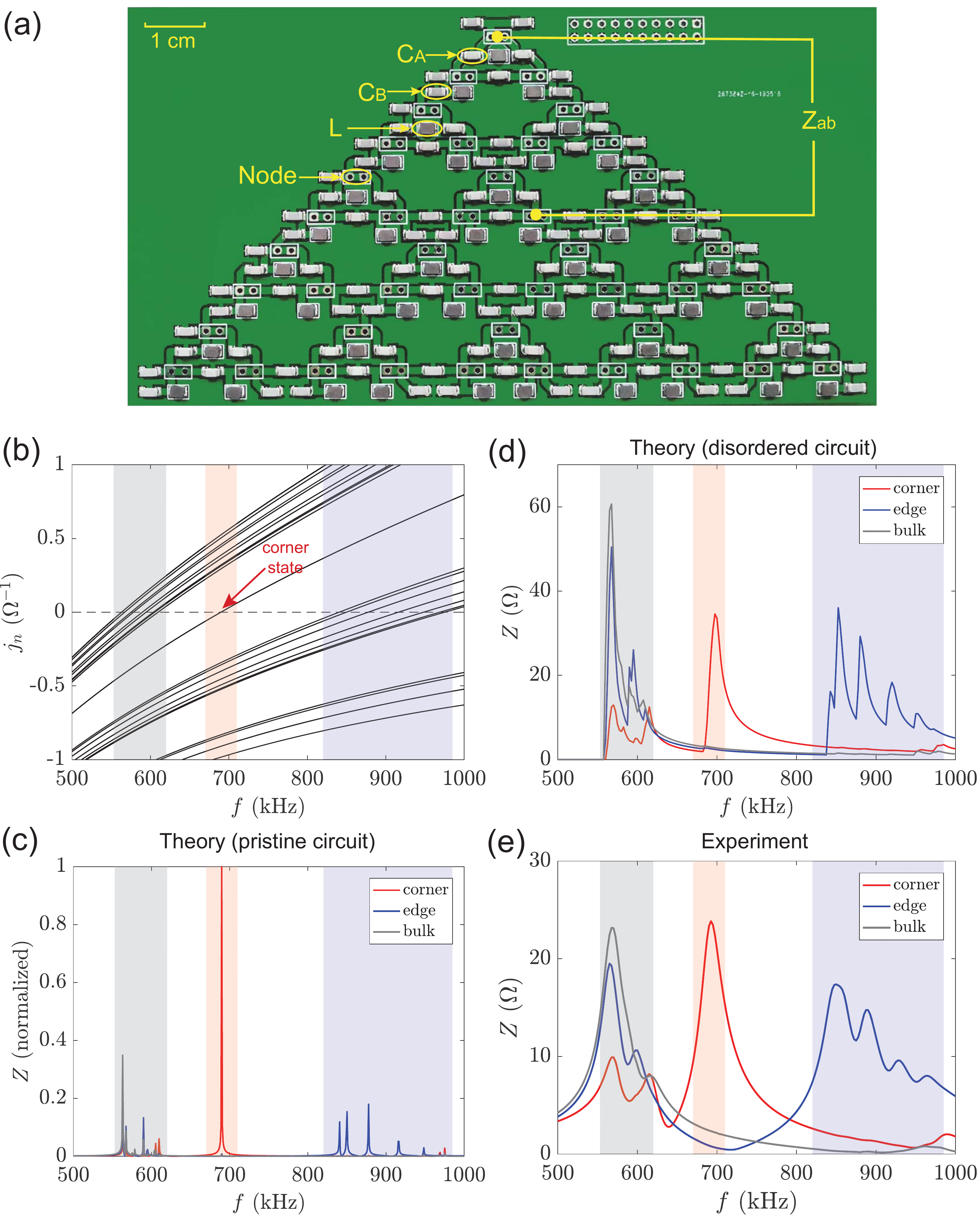}\\
  \caption{(a) Photograph of the layout of the experiment. (b) Theoretical spectrum of the circuit Laplacian $J(\omega)$ versus the driving frequency. Theoretical impedance in pristine (c) and disordered circuits (d). (e) Experimental results.}\label{Zab}
\end{figure}

\emph{Experimental observation.} We fabricate the real electrical-circuit on a Printed Circuit Board (PCB), as shown in Fig. \ref{Zab}(a) (see SM \cite{SM} for experimental details). We choose $C_A=22$ nF, $C_B=100$ nF, and $L=220$ nH, such that the ratio $C_A/C_B=0.22$ and the resonant frequency $f_c=\omega_c/2\pi=687$ kHz. Then, we measure the two-point impedance of the circuit:
\begin{equation}\label{Eq9}
Z_{ab}=\frac{V_a-V_b}{I_{ab}}=\sum_{n}\frac{|\psi_{n,a}-\psi_{n,b}|^2}{j_n},
\end{equation}
where $\psi_{n,a}-\psi_{n,b}$ is the amplitude difference between $a$ and $b$ nodes of the $n$th admittance mode (see SM \cite{SM} for the derivation). Due to the existence of zero-admittance corner states in a topologically non-trivial electric circuit, one should observe a strong peak at the resonant point $f=f_c$ by measuring the impedance spectrum between a corner node and another node [see Fig. \ref{Zab}(a)].

Figure \ref{Zab}(b) shows the theoretical spectrum of the circuit Laplacian $J(\omega)$ as a function of the driving frequency, where an isolated band localizes in the gap of spectrum, indicating the corner state. We compute the impendence between different nodes in both pristine and disordered circuits with numerical results plotted in Figs. \ref{Zab}(c) and \ref{Zab}(d), respectively. In the calculations, we considered the impedance between 13th and 15th $(Z_{13,15})$, 16th and 18th $(Z_{16,18})$, and 1st and 15th $(Z_{1,15})$ to quantify the signals from bulk, edge, and corner states, respectively; see Fig. \ref{ES}(a) for all numbered nodes. The disorder comes from the 2\% tolerance of circuit elements (see Table I in SM \cite{SM}). Theoretical results indeed show a strong peak emerging at the resonant point only from $Z_{1,15}$. Experimental measurement of the impedance in a frequency sweep manner is shown in Fig. \ref{Zab}(e), where a distinct peak of impedance appears at the resonant frequency, in excellent agreement with the disorder calculation.

\begin{figure}
  \centering
  \includegraphics[width=0.48\textwidth]{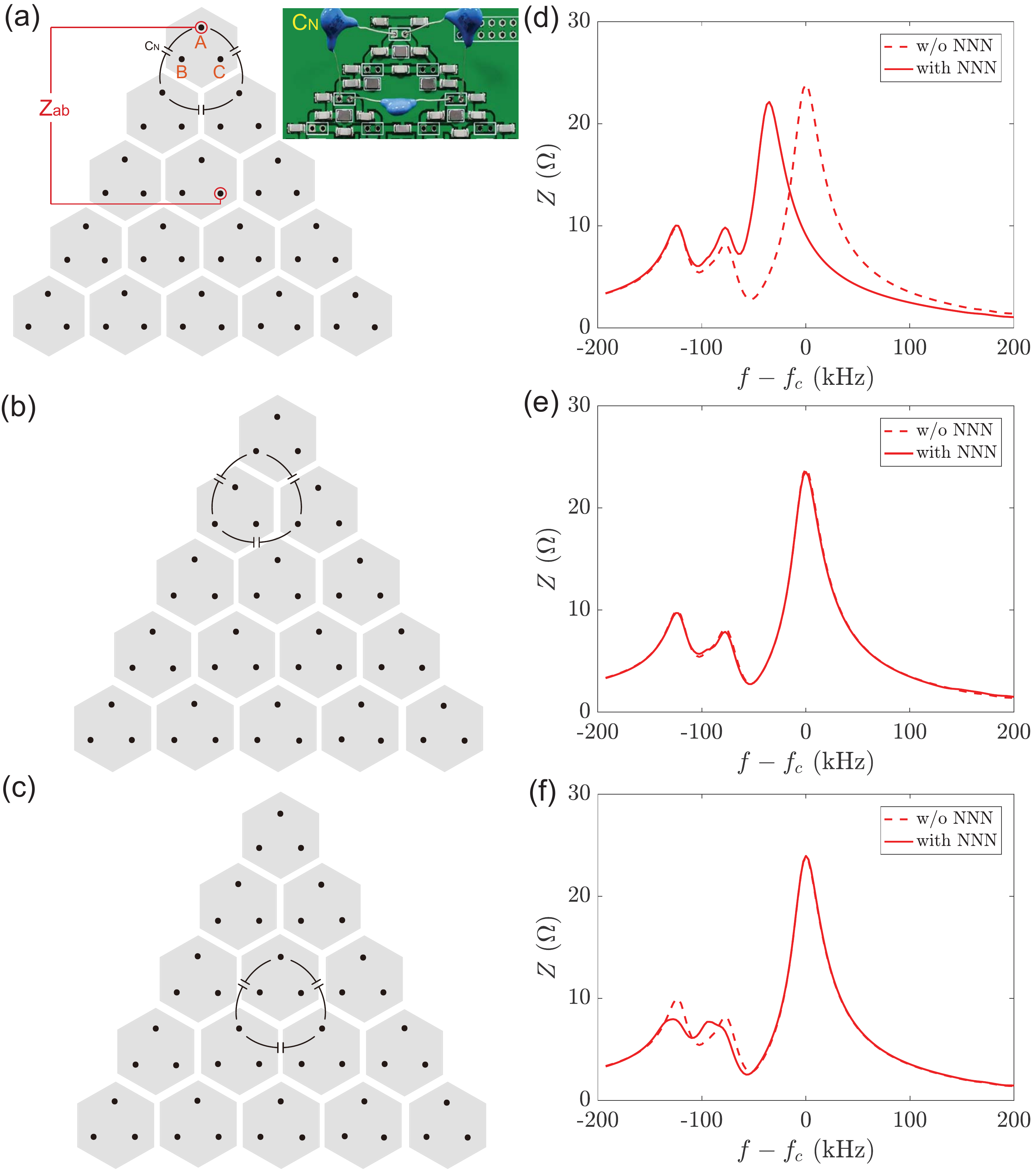}\\
  \caption{(a) Locally connecting three capacitors $C_N=10$ nF within sublattices A in the top corner. Inset shows the experiment setup. Introducing capacitors within sublattices B (b) and sublattices A in the bulk (c). (d)-(f) Experimental results for configurations considered in (a)-(c), respectively. The dashed curve is the measurement without $C_N$.}\label{NNN_hopping}
\end{figure}

\emph{Robustness of the zero mode.} To investigate the topological robustness of the zero mode, one should introduce the desired chiral-symmetry conserving and breaking disturbances, or the next-nearest-neighbour (NNN) hopping terms, which is realized by connecting extra capacitors $C_N$ within sublattices [see Fig. \ref{NNN_hopping}(a)]. We choose $C_N=10$ nF in our experiments.

\begin{figure}[htbp!]
  \centering
  \includegraphics[width=0.48\textwidth]{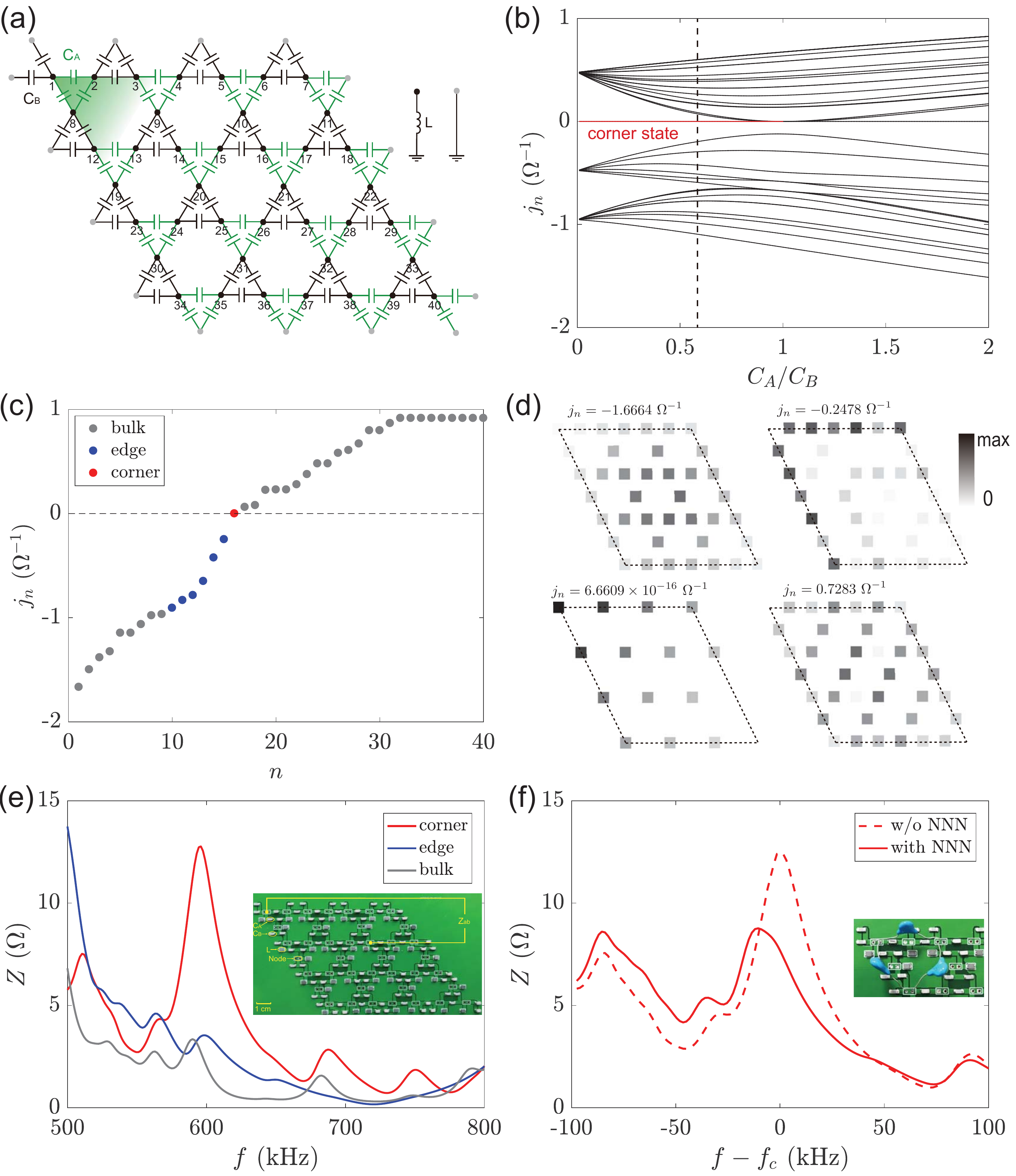}\\
  \caption{(a) Illustration of a rhombus breathing kagome circuit. (b) The admittance spectrum under resonant condition. The red line segment represents the emergence of zero modes in the top-left corner. (c) Admittances at $C_A/C_B$=0.68 [dashed black line in (b)]. (d) The spatial distribution of the bulk, edge, corner, and bulk modes, viewed from left to right and top to bottom, respectively. Experimental measurements of the impedances with (e) and without (f) the generalized chiral symmetry. Insets show the experiment setup.}\label{ESp}
\end{figure}
For convenience, we use A, B, and C to denote the three sublattices, as marked in Fig. \ref{NNN_hopping}(a). Three types of disturbances are introduced by connecting the capacitor $C_N$ within A sites in the corner, B sites in the edge, and A sites in the bulk, shown in Figs. \ref{NNN_hopping}(a), \ref{NNN_hopping}(b), and \ref{NNN_hopping}(c), respectively. The first one destroys the chiral symmetry of A sublattice in the top corner, while the latter two conserve it.

Experimental measurements of $Z_{1,15}$ for each consideration are depicted in Figs. \ref{NNN_hopping}(d), \ref{NNN_hopping}(e) and \ref{NNN_hopping}(f), respectively. The dashed curve is the experimental result without the extra capacitor (or the NNN hopping). For chiral-symmetry breaking case, we observe that the zero-mode peak at the resonant point is significantly suppressed and suffers a blue shift [see Fig. \ref{NNN_hopping}(d)]. Interestingly, the disturbance within B sites in the edge or A sites deep in the bulk leads to almost identical spectrum to that without it [see Figs. \ref{NNN_hopping}(e) and \ref{NNN_hopping}(f)]. It is worth mentioning that, in Fig. \ref{NNN_hopping}(c), the corner state is in principle affected but this effect is unnoticeable because the original state has almost no support in the region where the perturbation is implemented. Furthermore, we confirm that the zero modes of B and C sublattices survive in the bottom-left and bottom-right corners of the triangle circuit (not shown), respectively. Experimental results are fully consistent with theoretical calculations (see SM \cite{SM}).

\emph{Zero modes in rhombus-shape circuit.} To illustrate that the phase transition point is not at $C_A/C_B=1/2$, we study another kind of breathing kagome circuit (rhombus-shape), shown in Fig. \ref{ESp}(a). We calculate the admittance spectrum at the resonant frequency in Fig. \ref{ESp}(b). Now, it is rather clear that the zero-admittance mode always exists in one of two 60$^\circ$ corners, except for the critical point, i.e., $C_A/C_B=1$. For $C_A/C_B<1$ $(C_A/C_B>1)$, the zero mode appears in the top-left (bottom-right) corner, because the two opposite acute-angle corners are actually not identical \cite{Li2019_1}. By choosing $C_A/C_B=0.68$, we indeed identify an isolated corner state (zero mode) in the bulk gap, as shown in Fig. \ref{ESp}(c). Figure \ref{ESp}(d) plots the spatial distribution of representative eigenmodes for the bulk, edge, and corner states. Experimental measurements of the impedances ($Z_{1,16}$, $Z_{18,22}$, and $Z_{14,15}$ for corner, edge, and bulk states, respectively) are shown in Fig. \ref{ESp}(e), from which we observe a distinct $Z_{1,16}$ peak at the resonant frequency, in accordance with theoretical calculations (see SM \cite{SM}). Electric elements adopted in the rhombus experiments are $C_A=68$ nF, $C_B=100$ nF, and $L=220$ nH, such that the ratio $C_A/C_B=0.68$ and the resonant frequency $f_c=585$ kHz. When including the chiral-symmetry breaking disturbances, the zero mode again disappears, as shown in Fig. \ref{ESp}(f).
	
\emph{Conclusion.} To summarize, we presented the first experimental observation of topological corner states characterized by $\mathbb{Z}_3$ Berry phase in topolectrical circuits. We demonstrated theoretically and experimentally that the localized zero modes are protected by a generalized chiral symmetry of the tripartite lattice. By introducing capacitors within different sublattices, we examined the robustness of the zero modes against both chiral-symmetry conserving and breaking disturbances. We found a particular robustness of the zero mode as long as the generalized chiral symmetry is respected. Our work suggests that topolectrical circuits are ideal platform to realize exotic topological band model \cite{Yu2019,ChenR2019,ZhangZQ2019,Haenel2019}, which are challenging to implement in conventional condensed matter experiments.

\begin{acknowledgments}
\emph{Acknowledgments.} This work was supported by the National Natural Science Foundation of China (NSFC) (Grants No. 11604041 and 11704060), the National Key Research Development Program under Contract No. 2016YFA0300801, and the National Thousand-Young-Talent Program of China. Z.-X. Li acknowledges the financial support of the China Postdoctoral Science Foundation
(Grant No. 2019M663461) and the NSFC Grant No. 11904048.
\end{acknowledgments}

\end{document}